\documentclass[12pt]{article}

\usepackage{epsfig}

\newcommand{\mysection}{\setcounter{equation}{0}\section}

\def\beq{\begin{equation}}
\def\eeq{\end{equation}}
\def\beqa{\begin{eqnarray}}
\def\eeqa{\end{eqnarray}}

\newlength{\dinwidth} \newlength{\dinmargin}
\begin{document}

\begin{center}
{\Large \bf Soft-Gluon Corrections in FCNC Top-Quark Production via Anomalous Gluon Couplings}
\end{center}
\vspace{2mm}
\begin{center}
{\large Nikolaos Kidonakis$^a$ and Elwin Martin$^b$}\\
\vspace{2mm}
${}^a${\it Kennesaw State University,  Physics \#1202,\\
Kennesaw, GA 30144, USA} \\
\vspace{2mm}
${}^b${\it School of Physics, Georgia Institute of Technology, \\
Atlanta, GA 30332, USA}
\end{center}

\begin{abstract}
We present a calculation of soft-gluon corrections in FCNC top-quark production via anomalous $t$-$q$-$g$ couplings. The soft anomalous dimension matrix is explicitly calculated at one-loop accuracy. This calculation allows threshold resummation at next-to-leading-logarithm accuracy. We also derive expressions for the soft-gluon corrections at NLO and at NNLO. 
\end{abstract}

\mysection{Introduction}

Top quark production may provide insights into physics beyond the Standard Model, including tree-level flavor-changing neutral currents (FCNC). Because the top is very massive, new physics pertaining to Electroweak Symmetry Breaking will likely be most strongly coupled to the top quark. In several extensions of the Standard Model these tree-level FCNC processes involve anomalous couplings of the top quark, such as anomalous gluon couplings (see e.g. Refs. \cite{MT,HWYZ,TY96,HHWYZ}). Thus, FCNC top-quark production may provide a novel window into physics beyond the Standard Model.

Cross sections for such processes are usually given at leading order (LO). It is possible, however, to calculate a class of higher-order corrections from soft-gluon emission that are important near partonic threshold, i.e. when the energy of the incoming partons is just enough to produce a specified final state, with little energy left for additional radiation. Such corrections for FCNC top-quark processes with anomalous $t$-$q$-$V$ couplings, involving the top quark and a photon or Z boson, were studied in \cite{ABNK,NKAB,NKEM}.
It was found that the corrections at next-to-leading order (NLO) 
and next-to-next-to-leading order (NNLO) provide a significant enhancement of the cross section with a big decrease in the theoretical scale dependence. This is of course relevant for setting limits on $t$-$q$-$V$ anomalous couplings. 

Another class of anomalous couplings are those involving gluons, i.e. $t$-$u$-$g$ or $t$-$c$-$g$, where an up or charm quark interacts with a gluon and changes into a top quark. The Tevatron, HERA and, more recently, the LHC have searched for FCNC processes in the top-quark sector. ATLAS has set limits on $t$-$q$-$g$ couplings from 7 TeV \cite{ATLAS1} and 8 TeV \cite{ATLAS2} data (see also \cite{GSW}), and the latest values are  $\kappa_{tug}/\Lambda < 5.1 \times 10^{-3}$ TeV$^{-1}$ and $\kappa_{tcg}/\Lambda < 1.1 \times 10^{-2}$ TeV$^{-1}$ \cite{ATLAS2}.
In this paper we calculate the soft-gluon corrections for such FCNC processes. We keep the discussion general so that it can be applied to any specific model.

An effective FCNC Lagrangian for top quarks with gluons used in the LHC searches is given by:
\begin{equation}
\mathcal{L}_{\mbox{eff}} = g_s \sum_{q = u , c} \frac{\kappa_{tqg}}{\Lambda} \bar{t} \sigma^{\mu \nu} T^a \left( f^L_q P_L + f^R_q P_R \right) q G^a_{\mu \nu} + h.c.
\label{Leff}
\end{equation}
where $T^a$ are the Gell-Mann matrices, $\Lambda$ is a new-physics scale which we may take to be the top-quark mass, $\sigma^{\mu \nu} = \frac{i}{2} \left[ \gamma^\mu, \gamma^\nu \right]$, $P_{L,R}$ are the left and right projection operators, and $G^a_{\mu \nu}$ is the gauge-field tensor of the gluon.  

\begin{figure}
\begin{center}
\includegraphics[width=.3\linewidth]{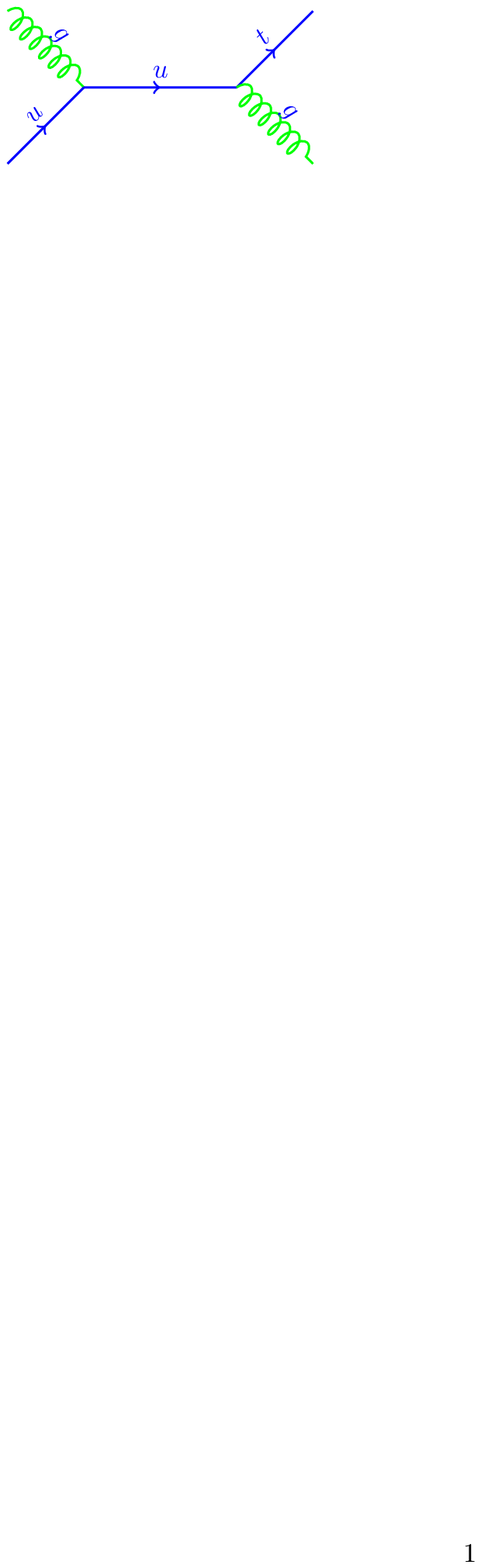}  \hspace{15mm}
\includegraphics[width=.3\linewidth]{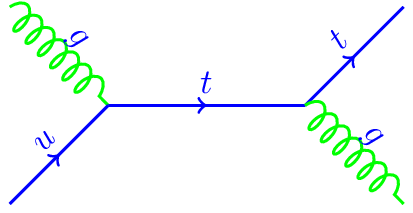}

\vspace{6mm}

\includegraphics[width=.3\linewidth]{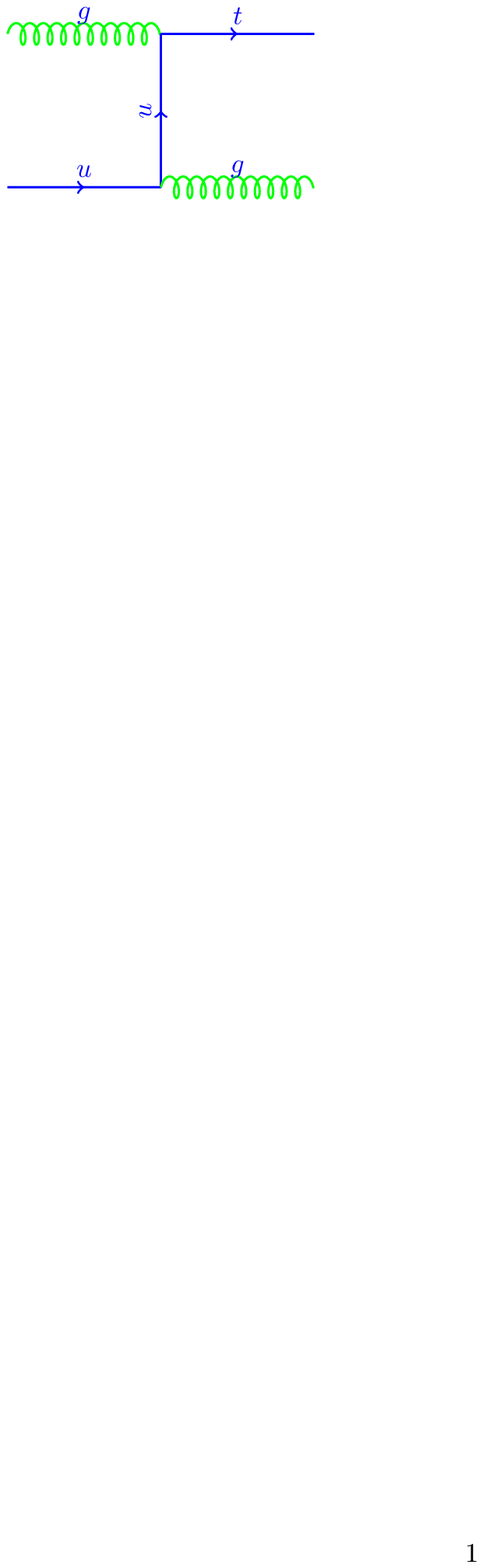} \hspace{15mm}
\includegraphics[width=.3\linewidth]{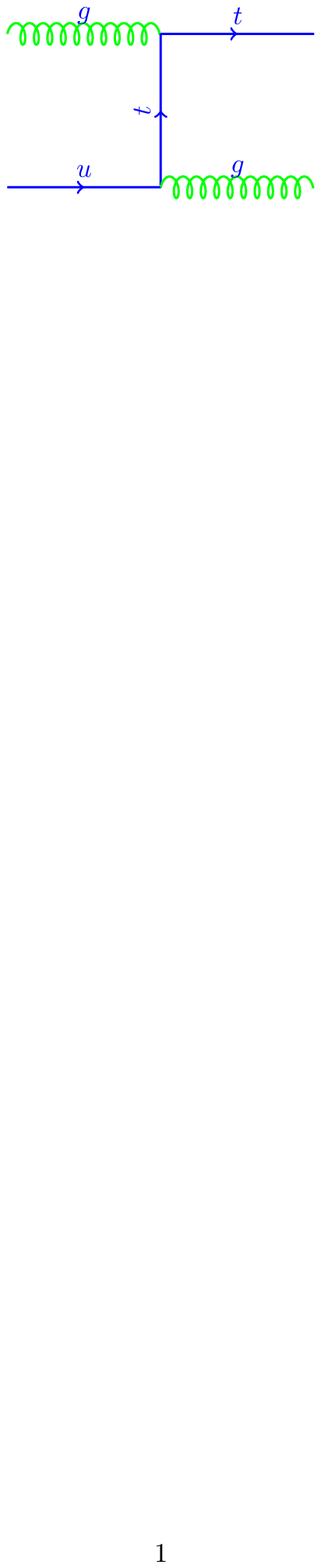}

\vspace{6mm}

\includegraphics[width=.3\linewidth]{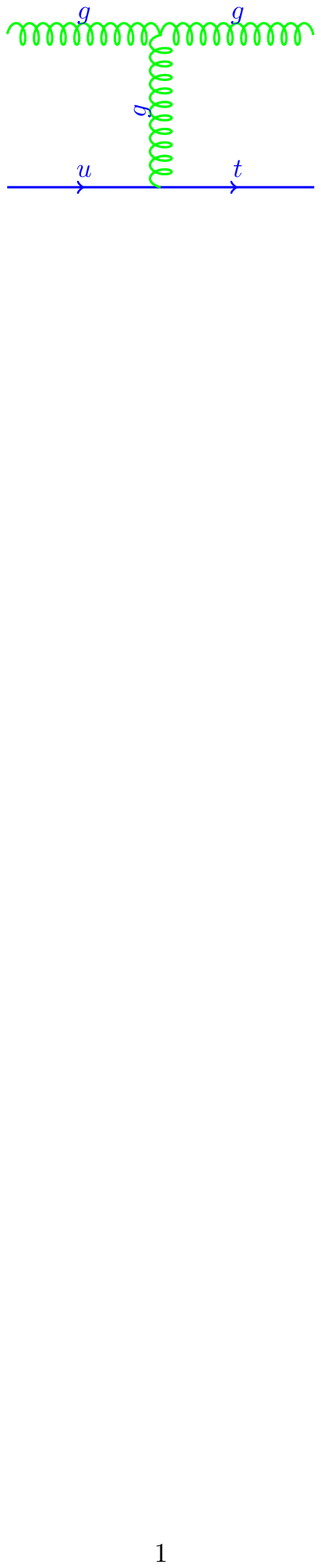} \hspace{15mm}
\includegraphics[width=.3\linewidth]{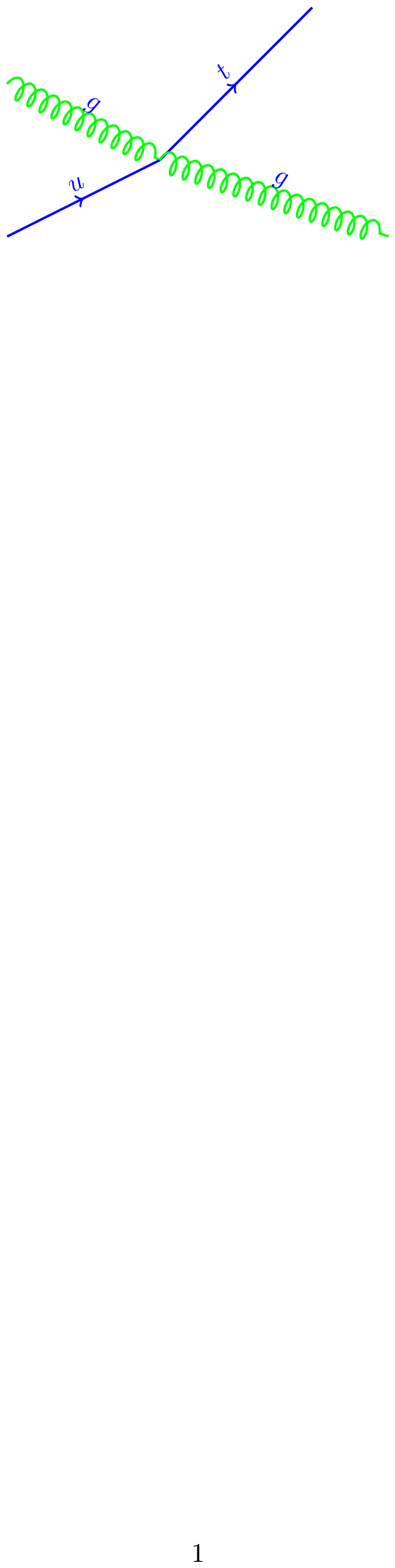}
\caption{The six diagrams above represent the various tree level processes for $gu \rightarrow tg$. 
There are two s-channel interactions and three t-channel interactions. The final diagram 
represents an atypical four-point interaction.}
\label{treelevel}
\end{center}
\end{figure}

For compactness we define $ \kappa_q = g_s \kappa_{tqg} / \Lambda$ and $\chi =  f^L_q P_L + f^R_q P_R $. For simplicity, in this paper we only consider $u$ contributions since the $c$ contributions are typically small in this regime. 
Of course the analytical results are the same if we have a charm quark instead of an up quark.
Explicitly, the compactified Lagrangian for the $t$-$u$-$g$ coupling is given by
\begin{equation}
\mathcal{L}_{\mbox{eff}} = \kappa_u \bar{t} \sigma^{\mu \nu} T^a \chi u G^a_{\mu \nu} \, .
\end{equation}
Recalling that $G^a_{\mu \nu} = \partial_{[ \mu} A^a_{ \nu ]} - i g_s  \left[A_\mu, A_\nu \right]^a $,
this expands to 
\begin{equation}
\mathcal{L}_{\mbox{eff}} = \kappa_u \bar{t} \sigma^{\mu \nu} T^a \chi u \partial_{[ \mu} A_{ \nu ]} - i g_s  \kappa_u \bar{t} \sigma^{\mu \nu} T^a \chi u \left[A_\mu, A_\nu \right]^a \, .
\label{Leff2}
\end{equation}
This gives rise to the tree-level diagrams shown in Figure \ref{treelevel}.

In the next section we discuss the resummation of soft-gluon corrections via factorization properties of the cross section in moment space. In Section 3 we calculate the relevant one-loop diagrams, and extract the UV pole structure of the results. In Section 4 we derive the soft anomalous dimension matrix that controls the resummation of soft-gluon emission, and present NLO and NNLO analytical expressions for the threshold corrections. We present some numerical results illustrating the size of the corrections in Section 5. We conclude in Section 6. Some details of our calculations are given in the Appendix.

\mysection{Factorization and Resummation}

The partonic processes for FCNC top production via anomalous gluon couplings are of the form
\beqa
g(p_1)\, + \, u\, (p_2) \rightarrow t(p_3)\, + g(p_4)\, + X  
\label{gutg}
\eeqa
where $X$ denotes any additional radiation beyond LO.
We define $s=(p_1+p_2)^2$, $t=(p_1-p_3)^2$, $u=(p_2-p_3)^2$. 
Also $s_4=s+t+u-m^2$, where $m$ is the top-quark mass. Thus, $s_4$ 
measures distance from partonic 
threshold, where there is no energy for additional radiation, but the top quark 
may have arbitrary momentum and is not restricted to be produced at rest. 
At partonic threshold $s_4=0$ and soft-gluon corrections appear in the perturbative expansion of the cross section in the form of plus distributions of logarithmic terms $[(\ln^k(s_4/m^2))/s_4]_+$, with power $k$ ranging from 0 to $2n-1$ for the $n$-th order corrections in the strong coupling, $\alpha_s$.
The plus distributions are 
defined by their integral with any smooth function $\phi$, such as
parton distributions, as 
\beqa
\int_0^{s_{4 \, max}} ds_4 \, \phi(s_4) \left[\frac{\ln^k(s_4/m^2)}
{s_4}\right]_{+} &=&
\int_0^{s_{4\, max}} ds_4 \frac{\ln^k(s_4/m^2)}{s_4} [\phi(s_4) - \phi(0)]
\nonumber \\ &&
{}+\frac{1}{k+1} \ln^{k+1}\left(\frac{s_{4\, max}}{m^2}\right) \phi(0) \, .
\label{splus}
\eeqa
These logarithmic terms can be formally resummed to all orders in the perturbative expansion \cite{NKGS}.

Resummation follows from the factorization properties of the 
cross section in moment space.
We define moments of the partonic cross section by  
${\sigma}(N)=\int (ds_4/s) \;  e^{-N s_4/s} {\sigma}(s_4)$, with $N$ the moment variable. 
Logarithms of $s_4$ in the physical cross section, $\sigma(s_4)$,  
give rise to logarithms of $N$ in the moment-space expression for 
the cross section, $\sigma(N)$, and the logarithms of $N$ appearing 
in $\sigma(N)$ exponentiate.

We write a factorized expression for the moment-space 
partonic scattering cross section in $n=4-\epsilon$ dimensions:
\beq
\sigma_{g u\rightarrow tgX}(N,\epsilon)= 
H_{IL}^{g u\rightarrow tgX} \left(\alpha_s(\mu)\right)\; 
S_{LI}^{g u \rightarrow tgX} 
\left(\frac{m}{N \mu},\alpha_s(\mu) \right)\;
\left(\prod J_{\rm in}\left (N,\mu,\epsilon \right) \right) 
J_{\rm out}\left (N,\mu,\epsilon \right)
\label{factsigma}
\eeq 
with $\mu$ the scale, and where we denote incoming and outgoing parton 
jet functions by $J_{\rm in}$ and $J_{\rm out}$, respectively, which describe 
soft and collinear emission from the external partons. 
The hard-scattering terms $H_{IL}^{gu\rightarrow tgX}$ involve contributions 
from the amplitude of the process, $h$, and its complex conjugate, $h^*$, 
in the form $H_{IL}=h_L^*\, h_I$, where $I$, $L$ are color indices. 
Also, $S_{LI}^{gu\rightarrow tgX}$ is the soft gluon function for 
non-collinear soft-gluon emission \cite{NKGS,KOS};
it represents the coupling of soft gluons to the
partons in the scattering with color tensors $c_I$, $c_L$. 
Both $H_{IL}$ and $S_{LI}$ are process dependent and 
they are $3 \times 3$ matrices in the space of color exchanges in the 
partonic scattering for the process $gu \rightarrow tg$. Because of the non-trivial color structure of this process the calculation is more complicated than for the cases of $gu \rightarrow tZ$ 
and $gu \rightarrow t\gamma$ studied in \cite{NKAB} where the soft anomalous 
dimension was simply a function, i.e. a $1\times 1$ matrix.

The requirement that the product of the factors in Eq. (\ref{factsigma}) be
independent of the gauge and the factorization scale results
in the exponentiation of logarithms of $N$. 
The soft matrix $S_{LI}$ requires renormalization and   
its $N$-dependence can then be resummed via renormalization group evolution 
(RGE) \cite{NKGS}. We have 
\beq
S^b_{LI}=Z^\dagger_{S\, LC} \, S_{CD} \, Z_{S\, DI}
\eeq
where $S^b$ is the unrenormalized quantity,
and $Z_S$ is a $3 \times 3$ matrix of renormalization constants.

Thus  $S_{LI}$ satisfies the renormalization group equation
\beq
\left(\mu \frac{\partial}{\partial \mu}
+\beta(g_s, \epsilon)\frac{\partial}{\partial g_s}\right)\,S_{LI}
=-\Gamma^\dagger_{S\, LC} \, S_{CI}-S_{LD} \, \Gamma_{S\, DI}
\eeq
where $g_s^2=4\pi\alpha_s$; 
$\beta(g_s, \epsilon)=-g_s \epsilon/2 + \beta(g_s)$ 
where $\beta(g_s)$ is the QCD beta function
\beq
\beta(g_s) \equiv \mu \frac{d g_s}{d \mu}
=-\beta_0 \frac{g_s^3}{16 \pi^2}+\cdots \, ,
\eeq
with $\beta_0=(11C_A-2n_f)/3$, $C_A=N_c$, $N_c=3$ the number of colors, 
and $n_f=5$ the number of light quark flavors;  
and $\Gamma_S$ with matrix elements
\beq
\Gamma_{S\, DI}=\left(\frac{d}{d\ln\mu} Z_{S\, DK}\right) Z^{-1}_{S\, KI}
=\beta(g_s, \epsilon)  \frac{\partial Z_{S\, DK}}{\partial g_s} Z^{-1}_{S\, KI}
\eeq
is the soft anomalous dimension matrix 
that controls the evolution of the soft function $S$. 
In dimensional regularization $Z_S$ has $1/\epsilon$ poles. 
Expanding $Z_S$ in powers of the strong coupling, 
\beq 
Z_{S \, DK}=\delta_{DK}+\frac{\alpha_s}{\pi}Z^{(1)}_{S \, DK}+{\cal O}(\alpha_s^2) \, ,
\eeq 
and since $Z^{(1)}_S$ has a $1/\epsilon$ pole while $\beta(g_s, \epsilon)$  
includes a $-g_s \epsilon/2$ term in dimensional regularization, 
we find that $\Gamma_S$ is given at one loop simply by minus the residue of 
$Z_S$.
The soft anomalous dimension $\Gamma_S$ is a $3 \times 3$ matrix in 
color space and a function of the kinematical invariants $s$, $t$, $u$.

The resummed cross section in moment space  
follows from the RGE of all the functions in the factorized cross section, 
Eq. (\ref{factsigma}), and can be written in the form:
\beqa
{\sigma}_{gu\rightarrow tgX}^{\rm resummed}(N) &=&
\exp\left[ \sum_{i=g,u} E_i(N_i)\right] \, \exp\left[E'_g(N')\right]\;
\nonumber\\ && \times \,
{\rm tr} \left\{H^{gu\rightarrow tgX}\left(\alpha_s(\sqrt{s})\right)
\exp \left[\int_{\sqrt{s}}^{{\sqrt{s}}/{\tilde N'}}
\frac{d\mu}{\mu} \;
\Gamma_S^{\dagger \, gu\rightarrow tgX}\left(\alpha_s(\mu)\right)\right] \right.
\nonumber\\ && \left. \times \,
S^{gu\rightarrow tgX} \left(\alpha_s\left(\frac{\sqrt{s}}{\tilde N'}\right)
\right) \;
\exp \left[\int_{\sqrt{s}}^{{\sqrt{s}}/{\tilde N'}}
\frac{d\mu}{\mu}\; \Gamma_S^{gu\rightarrow tgX}
\left(\alpha_s(\mu)\right)\right] \right\}
\label{ressigma}
\eeqa
where the first two exponents resum soft and collinear radiation from the incoming gluon and quark and from the outgoing gluon, respectively, and have well-known expressions \cite{GS87,CT89}. Here $N_g=N(-u/m^2)$, $N_q=N(-t/m^2)$, $N'=N(s/m^2)$ and ${\tilde N}=N e^{\gamma_E}$ with $\gamma_E$ the Euler constant.
The trace is taken of the product of the color-space matrices $H$, $S$, 
and exponents of $\Gamma_S$ and its Hermitian conjugate, $\Gamma_S^{\dagger}$.
Noncollinear soft gluon emission is controlled by the soft anomalous dimension 
$\Gamma_S$, which has the perturbative expansion 
\beq
\Gamma_S^{gu\rightarrow tgX}=\frac{\alpha_s}{\pi} \Gamma_S^{(1)}
+\frac{\alpha_s^2}{\pi^2} \Gamma_S^{(2)}+\cdots
\eeq
We also write the expansions $H^{gu\rightarrow tgX}=\alpha_s^2 H^{(0)}+(\alpha_s^3/\pi)H^{(1)}\cdots$ and $S^{gu\rightarrow tgX}=S^{(0)}+(\alpha_s/\pi)S^{(1)}+\cdots$.

We determine $\Gamma_S$ from the coefficients of ultraviolet poles 
in dimensionally regularized eikonal diagrams.
The determination of $\Gamma_S^{(1)}$ is needed for next-to-leading logarithm 
(NLL) resummation and it requires one-loop calculations in 
the eikonal approximation.

\mysection{UV poles of one-loop diagrams}

In this section we calculate the one-loop diagrams needed for the determination
of the one-loop soft anomalous dimension matrix, $\Gamma_S^{(1)}$.

For the partonic process 
\begin{equation}
g(p_1,a) + u(p_2,b) \to t(p_3,i) + g(p_4,j)
\end{equation}
with momenta $p$ and color labels $a,b$ for the incoming partons and $i,j$ for the outgoing ones, we choose the color basis 
\begin{equation}
c_1 = \delta_{b i} \delta_{a j}, ~~ c_2 = d^{a j c} T^c_{i b}, ~~ c_3 = i f^{a j c} T^c_{i b} \, .
\label{colorbasis}
\end{equation}

We detail some of the results necessary to calculate the soft-anomalous dimension matrix, $\Gamma_S$, at one loop.
For the process that we are considering there are six one-loop vertex corrections and a self-energy correction to the massive top quark. 
The poles from the one-loop vertex calculations are extracted for each diagram,
including the self-energy contributions.

\begin{figure}
\begin{center}
\includegraphics[width=.3\linewidth]{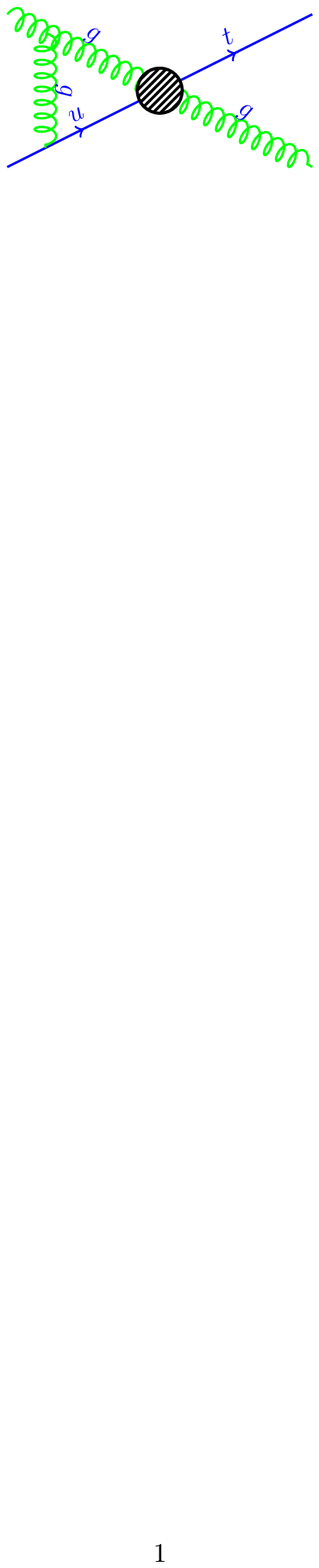}  \hspace{15mm}
\includegraphics[width=.3\linewidth]{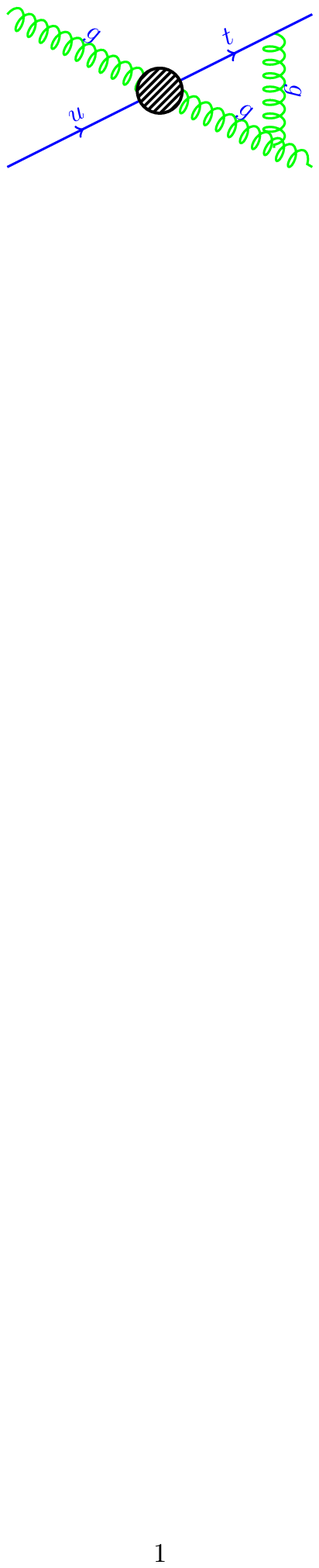}
\caption{One-loop eikonal diagrams 1 (left) and 2 (right).}
\label{diag12}
\end{center}
\end{figure}

\begin{figure}
\begin{center}
\includegraphics[width=.3\linewidth]{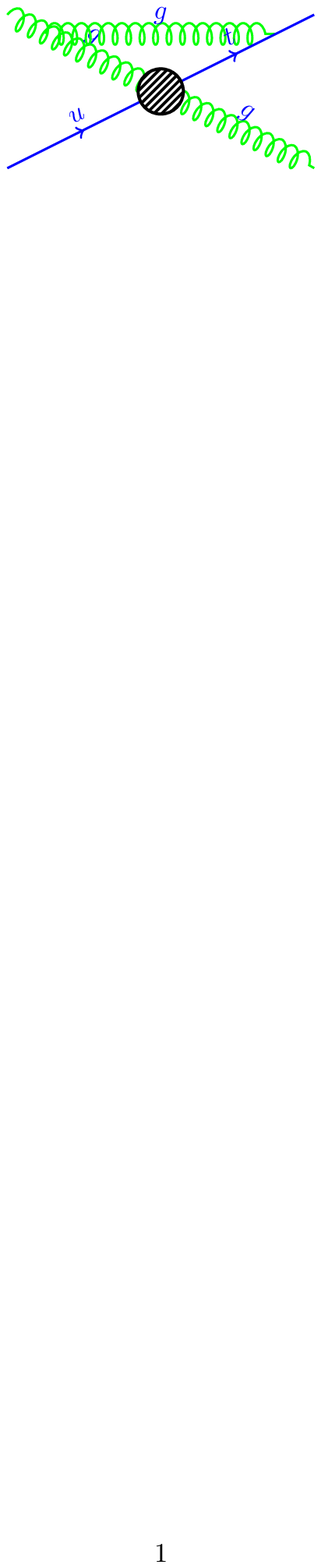}  \hspace{15mm}
\includegraphics[width=.3\linewidth]{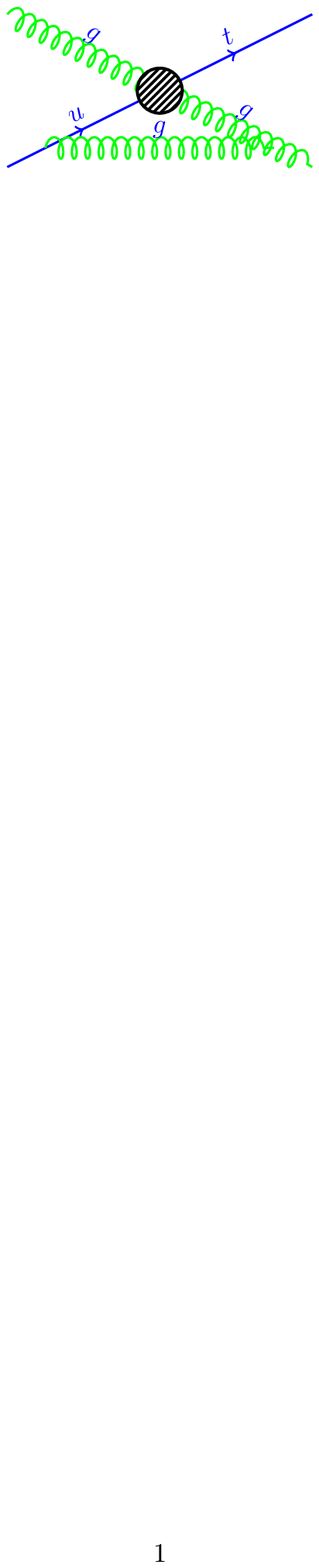}
\caption{One-loop eikonal diagrams 3 (left) and 4 (right).}
\label{diag34}
\end{center}
\end{figure}

\begin{figure}
\begin{center}
\includegraphics[width=.3\linewidth]{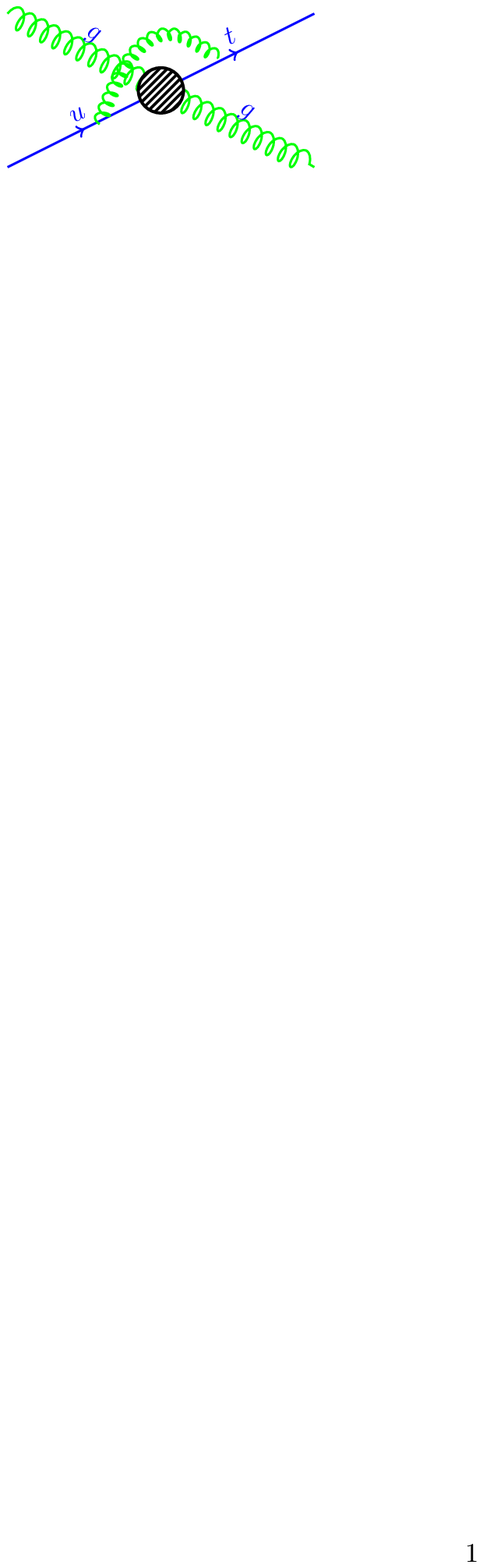}  \hspace{15mm}
\includegraphics[width=.3\linewidth]{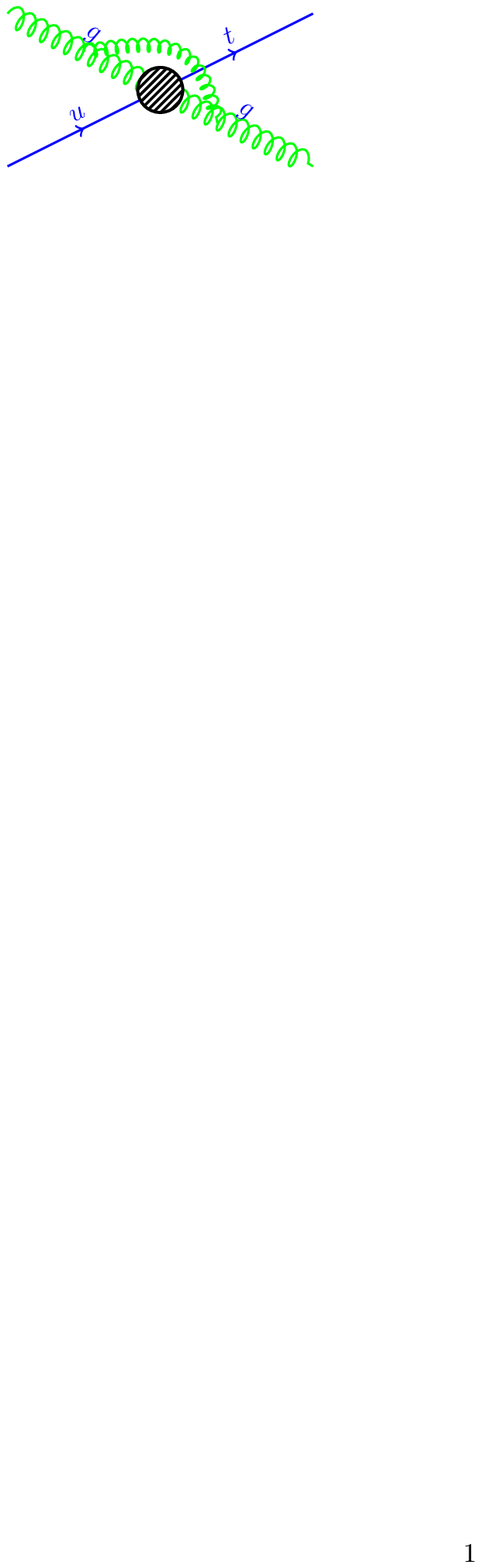}
\caption{One-loop eikonal diagrams 5 (left) and 6 (right).}
\label{diag56}
\end{center}
\end{figure}

The UV divergent contributions to 
each $S_{LI}$ from one-loop vertex corrections are shown in the eikonal diagrams of Figs. \ref{diag12}, \ref{diag34}, \ref{diag56}. The dark blobs in these diagrams denote the color basis tensors.
The counterterms for $S$ are the ultraviolet divergent
coefficients times our basis color tensors, $c_I$.

We perform our calculations in momentum space and in Feynman gauge.
We use the eikonal approximation, where the usual Feynman rules are simplified by letting the gluon momentum approach zero (see the Appendix).
Let $v^{\mu}$ be a velocity vector proportional to $p^{\mu}$; we write 
$p^\mu = \sqrt{s/2} \, v^\mu $. Then the eikonal quark-gluon vertex is  $v^{\mu}/(v\cdot k+i\epsilon)$ (with additional plus and minus signs) and similarly for the eikonal three-gluon vertex (with additional factors of $i$) as detailed in the Appendix.


We denote the integral describing diagram 1 as $I_1$, that for diagram 2 as $I_2$, etc. Thus we have the expressions
\beqa
I_{1} &=& g_s^2 \int \frac{d^n k}{(2 \pi)^n}  \frac{(-i v_1^\mu)}{(-v_1 \cdot k +i\epsilon)}  \frac{(-i) g_{\mu \nu}}{k^2}   \frac{v_2^\nu}{(v_2 \cdot k+i\epsilon)}=i \, I(v_1,v_2)
\nonumber \\
I_{2} &=& g_s^2 \int \frac{d^n k}{(2 \pi)^n}   \frac{v_3^\mu}{(v_3 \cdot k +i\epsilon)}  \frac{(-i) g_{\mu \nu}}{k^2}  \frac{i v_4^\nu}{(-v_4 \cdot k+i\epsilon)}=-i \, I(v_3, v_4)
\nonumber \\
I_{3} &=& g_s^2 \int \frac{d^n k}{(2 \pi)^n}  \frac{i v_1^\mu}{(-v_1 \cdot k +i\epsilon)}  \frac{(-i) g_{\mu \nu}}{k^2}  \frac{v^\nu_3}{(-v_3 \cdot k+i\epsilon)}=i\, I(v_1,v_3)
\nonumber \\
I_{4} &=& g_s^2 \int \frac{d^n k}{(2 \pi)^n}  \frac{v_2^\mu}{(-v_2 \cdot k +i\epsilon)}  \frac{(-i) g_{\mu \nu}}{k^2}  \frac{(-iv_4^\nu)}{(-v_4 \cdot k+i\epsilon)}=-i \, I(v_2,v_4)
\nonumber \\
I_{5} &=& g_s^2 \int \frac{d^n k}{(2 \pi)^n}  \frac{ v_2^\mu}{(-v_2 \cdot k +i\epsilon)}  \frac{(-i) g_{\mu \nu}}{k^2}  \frac{v^\nu_3}{(-v_3 \cdot k+i\epsilon)}=I(v_2,v_3)
\nonumber \\
I_{6} &=& g_s^2 \int \frac{d^n k}{(2 \pi)^n}  \frac{i v_1^\mu}{(-v_1 \cdot k +i\epsilon)}  \frac{(-i) g_{\mu \nu}}{k^2}  \frac{iv^\nu_4}{(-v_4 \cdot k+i\epsilon)}=-I(v_1,v_4)
\eeqa
where we have used the definition
\beq
I(v_l,v_m) = g_s^2 \int \frac{d^n k}{(2 \pi)^n}  \frac{ v_l^\mu}{(v_l \cdot k +i\epsilon)}  \frac{(-i)g_{\mu \nu}}{k^2}  \frac{v^\nu_m}{(v_m \cdot k+i\epsilon)} \, .
\eeq

Using Feynman parameterization and relations for $n$-dimensional integrals \cite{NKtW}, we can separate and determine  the UV pole of $I(v_l,v_m)$ which is:
\beq
I^{\rm UV}(v_l,v_m) = - \frac{1}{\epsilon}\frac{\alpha_s}{\pi} \ln \left(\frac{2v_l \cdot v_m} {\sqrt{v_l^2 v_m^2}}\right) \, .
\eeq
We can apply this result to find the UV poles of all six diagrams. However, since $v^2=0$ for massless quarks, we need to add to the above result terms $-(\alpha_s/\pi\epsilon)\ln\sqrt{v^2/2}$ from the eikonal lines for all massless quarks in each diagram to get the final expressions for the UV poles of all the above six diagrams. We denote each such final expression as $I_1^{\rm UV}$, $I_2^{\rm UV}$, etc.
Thus, we find explicitly
\beqa 
I_1^{\rm UV}&=&-\frac{i}{\epsilon} \frac{\alpha_s}{\pi} \ln(v_1 \cdot v_2)
\nonumber \\
I_2^{\rm UV}&=&\frac{i}{\epsilon} \frac{\alpha_s}{\pi} 
\ln\left(\frac{\sqrt{2} v_3 \cdot v_4}{\sqrt{v_3^2}}\right)
\nonumber \\
I_3^{\rm UV}&=&-\frac{i}{\epsilon} \frac{\alpha_s}{\pi} 
\ln\left(\frac{\sqrt{2} v_1 \cdot v_3}{\sqrt{v_3^2}}\right)
\nonumber \\
I_4^{\rm UV}&=&\frac{i}{\epsilon} \frac{\alpha_s}{\pi} \ln(v_2 \cdot v_4)
\nonumber \\
I_5^{\rm UV}&=&-\frac{1}{\epsilon} \frac{\alpha_s}{\pi} 
\ln\left(\frac{\sqrt{2} v_2 \cdot v_3}{\sqrt{v_3^2}}\right)
\nonumber \\
I_6^{\rm UV}&=&\frac{1}{\epsilon} \frac{\alpha_s}{\pi} \ln(v_1 \cdot v_4)
\eeqa

Finally we have to consider the self-energy corrections to the massive top-quark at one-loop. The result is well known and is given by
\begin{equation}
I^{\rm UV}_{\rm SE} = \frac{\alpha_s}{\pi} \frac{1}{2\epsilon}
\end{equation}

As mentioned above, the counterterms for $S$ are the ultraviolet divergent
coefficients times our basis color tensors, $c_I$. We have
\begin{eqnarray}
S_1&=&c_1 Z_{11} + c_2 Z_{21} + c_3 Z_{31},
\nonumber\\
S_2&=&c_1 Z_{12} + c_2 Z_{22} + c_3 Z_{32},
\nonumber\\
S_3&=&c_1 Z_{13} + c_2 Z_{23} + c_3 Z_{33}.
\label{ZLI}
\end{eqnarray}
The first line in Eq. (\ref{ZLI}) is for corrections to the color tensor $c_1$, 
the second line is for $c_2$ and the third for $c_3$.

Using the above results and the expressions for the color factors in the Appendix we find
\beqa
Z_{11}&=&C_F I_5^{\rm UV}-C_A I_6^{\rm UV}+C_F I_{\rm SE}^{\rm UV}
\nonumber \\
Z_{21}&=&0
\nonumber \\
Z_{31}&=&i(I_1^{\rm UV}-I_2^{\rm UV}-I_3^{\rm UV}+I_4^{\rm UV})
\nonumber \\
Z_{12}&=&0
\nonumber \\
Z_{22}&=&-i\frac{N_c}{4} (I_1^{\rm UV}-I_2^{\rm UV}+I_3^{\rm UV}-I_4^{\rm UV})+\left(C_F-\frac{C_A}{2}\right)I_5^{\rm UV}-\frac{C_A}{2}I_6^{\rm UV}+C_F I_{\rm SE}^{\rm UV}
\nonumber \\
Z_{32}&=&i\left(\frac{N_c}{4}-\frac{1}{N_c}\right)(I_1^{\rm UV}-I_2^{\rm UV}-I_3^{\rm UV}+I_4^{\rm UV})
\nonumber \\
Z_{13}&=&\frac{i}{2}(I_1^{\rm UV}-I_2^{\rm UV}-I_3^{\rm UV}+I_4^{\rm UV})
\nonumber \\
Z_{23}&=&i\frac{N_c}{4}(I_1^{\rm UV}-I_2^{\rm UV}-I_3^{\rm UV}+I_4^{\rm UV})
\nonumber \\
Z_{33}&=&Z_{22}
\label{Zmatrix}
\eeqa
The elements of the soft anomalous dimension matrix can then be simply read from the above equation by dropping overall $-1/\epsilon$ factors.

\mysection{Soft Anomalous Dimension  Matrix and soft-gluon corrections at NLO and NNLO}

Using the above results we are now ready to write the expression for the soft anomalous dimension matrix at one loop. 
First we express the dot products between the $v^{\mu}$ vectors in terms of 
the kinematical variables $s$, $t$, $u$.
This gives:
\beqa
&& v_1 \cdot v_2 = 1 \, , \quad \quad v_1 \cdot v_4 = - \frac{u}{s} \, , 
\quad \quad v_3 \cdot v_4 =  \frac{s- m^2}{s} \, , 
\nonumber \\ &&
v_2 \cdot v_3 = \frac{ m^2 - u}{s} \, , \quad \quad v_1 \cdot v_3 =  \frac{m^2 -t}{s} \, , \quad \quad v_2 \cdot v_4 = - \frac{t}{s} \, . 
\eeqa
Also, $\sqrt{v_3^2} = \sqrt{2/s} \, m $.

The explicit result for the one-loop soft anomalous dimension matrix 
in terms of these kinematic variables is 
\beq
\Gamma_S^{(1)} =  \left[\begin{array}{ccc}
\Gamma_{11} & \Gamma_{12} & \Gamma_{13} \\
\Gamma_{21} & \Gamma_{22} & \Gamma_{23} \\
\Gamma_{31} & \Gamma_{32}  & \Gamma_{33}
\end{array}
\right] 
\eeq
where 
\beqa
\Gamma_{11} &=& C_F \left(  \ln \left( \frac{m^2-u}{m\sqrt{s}} \right) - \frac{1}{2} \right) + C_A \ln \left( \frac{-u}{s} \right)  
\nonumber \\
\Gamma_{12} &=& \Gamma_{21} = 0 
\nonumber \\
\Gamma_{31} &=& \ln \left( \frac{t(t-m^2)}{s(s-m^2)} \right)  
\nonumber \\
\Gamma_{32} &=& \frac{(N_c^2-4)}{4N_c} \Gamma_{31} 
\nonumber \\
\Gamma_{22} &=& \Gamma_{33} =  C_F \left(  \ln \left( \frac{m^2-u}{m\sqrt{s}} \right) - \frac{1}{2} \right) +  \frac{C_A}{4} \ln \left( \frac{t u^2 (s-m^2)(t-m^2)}{(m^2-u)^2 s^3} \right) 
\nonumber \\
\Gamma_{23} &=& \frac{C_A}{4} \Gamma_{31} 
\nonumber \\
\Gamma_{13} &=& \frac{1}{2}  \Gamma_{31} 
\eeqa

From the moment-space expression for the resummed cross section, Eq. (\ref{ressigma}), we can derive upon inversion to momentum space the soft-gluon corrections in the perturbative expansion of the physical cross section in the strong coupling. Here we present explicitly the NLO and NNLO soft-gluon corrections. 

The NLO soft-gluon corrections are 
\beqa
\frac{d\sigma^{(1)}_{gu\rightarrow tg}}{dt \, du} &=& F^B_{gu\rightarrow tg} 
\frac{\alpha_s(\mu_R)}{\pi}
\left\{c_3^{gu\rightarrow tg} \left[\frac{\ln(s_4/m^2)}{s_4}\right]_+ 
+c_2^{gu\rightarrow tg}  \left[\frac{1}{s_4}\right]_+ 
+c_1^{gu\rightarrow tg} \, \delta(s_4) \right\}
\nonumber \\ &&
{}+\frac{\alpha_s^3(\mu_R)}{\pi} \, A^{gu\rightarrow tg}  
\left[\frac{1}{s_4}\right]_+
\eeqa
where $F^B_{gu\rightarrow tg}$ is the Born term, and the coefficients are given by $c_3^{gu\rightarrow tg}=2C_F+C_A$, 
\beq
c_2^{gu\rightarrow tg}=T_2^{gu\rightarrow tg}
-(C_F+C_A) \ln\left(\frac{\mu_F^2}{m^2}\right)
\eeq
\beq 
T_2^{gu\rightarrow tg}=-2 C_F \ln\left(\frac{-t}{s}\right)-2 C_A \ln\left(\frac{-u}{s}\right)-\frac{\beta_0}{4}+C_F \ln\left(\frac{m^2}{s}\right)
\eeq
\beq
c_1^{gu\rightarrow tg}=\left[C_F\, \ln\left(\frac{-t}{m^2}\right) 
+C_A \, \ln\left(\frac{-u}{m^2}\right) 
-\frac{3}{4}C_F-\frac{\beta_0}{4}\right]\ln\left(\frac{\mu_F^2}{m^2}\right)
+\frac{\beta_0}{2} \ln\left(\frac{\mu_R^2}{m^2}\right) 
\eeq
and
\beq
A^{gu\rightarrow tg}={\rm tr} \left(H^{(0)} \Gamma_S^{(1)\,\dagger} S^{(0)}
+H^{(0)} S^{(0)} \Gamma_S^{(1)}\right) \, .
\label{Ac}
\eeq
Note that in the coefficient $c_1^{gu\rightarrow tg}$ we have only provided scale-logarithm terms, i.e. terms with the logarithms of the factorization scale, $\mu_F$, and the renormalization scale, $\mu_R$.
A complete NLO calculation would be required to determine the scale-independent terms in $c_1^{gu\rightarrow tg}$.

The lowest-order soft-matrix elements, used in Eq. (\ref{Ac}), 
are given by $S^{(0)}_{LI}=c_L^* \, c_I$, so  
$S^{(0)}$ is diagonal and given explicitly by 
\beq
S^{(0)} =  \left[\begin{array}{ccc}
N_c(N_c^2-1) & 0 & 0 \\
0 & C_F(N_c^2-4) & 0 \\
0 & 0  & N_c(N_c^2-1)/2 
\end{array} 
\right] \, .
\eeq

The lowest-order hard matrix $H^{(0)}$, used in Eq. (\ref{Ac}), 
depends on the specific form of the Lagrangian for the process and can be simply derived from a color decomposition of the Born cross section, since $F^B={\rm tr}(H^{(0)} S^{(0)})$. We have derived such results for the specific Lagrangian of Eq. (\ref{Leff2}), which we present in the next section, but we note that the expressions would differ for other specifications for the Lagrangian. However all the other results in this section are independent of the details of the anomalous coupling.

With the one-loop NLL calculation of the soft anomalous dimension matrix we have been able to determine the full coefficients of all logarithmic terms at NLO. 
At NNLO the highest power of logarithm is cubic. With our current NLL calculation we can only fully determine the coefficients of the cubic and square powers of the logarithms of $s_4$ in the plus distributions at NNLO.  We can further calculate the $\zeta$ terms and all the scale-logarithm terms of the single-power logarithms of $s_4$; and the $\zeta$ terms and squared scale-logarithm terms of the zeroth-power of the logarithms of $s_4$. The $\zeta$ terms (i.e. the terms involving the constants $\zeta_2=\pi^2/6$ and $\zeta_3=1.2020569\cdots$) arise from the inversion from moment to momentum space. We thus find the NNLO expression for the soft-gluon corrections  
\beqa
\frac{d\sigma^{(2)}_{gu\rightarrow tg}}{dt \, du}&=&
 F^B_{gu\rightarrow tg} \frac{\alpha_s^2(\mu_R)}{\pi^2}
\frac{1}{2} (c_3^{gu\rightarrow tg})^2\, 
\left[\frac{\ln^3(s_4/m^2)}{s_4}\right]_+ 
\nonumber \\ && \hspace{-15mm} 
{}+F^B_{gu\rightarrow tg} \frac{\alpha_s^2(\mu_R)}{\pi^2} 
\left[\frac{3}{2}c_3^{gu\rightarrow tg} c_2^{gu\rightarrow tg}-\frac{\beta_0}{4} 
c_3^{gu\rightarrow tg} +\frac{\beta_0}{8} C_A\right]  
\left[\frac{\ln^2(s_4/m^2)}{s_4}\right]_+ 
\nonumber \\ && \hspace{-15mm}
{}+\frac{\alpha_s^4(\mu_R)}{\pi^2} 
\frac{3}{2} c_3^{gu\rightarrow tg} A^{gu\rightarrow tg} \,  
\left[\frac{\ln^2(s_4/m^2)}{s_4}\right]_+
\nonumber \\ && \hspace{-15mm}
{}+F^B_{gu\rightarrow tg} \frac{\alpha_s^2(\mu_R)}{\pi^2} 
\left[c_3^{gu\rightarrow tg} c_1^{gu\rightarrow tg}
-2 T_2^{gu\rightarrow tg} (C_F+C_A) \ln\left(\frac{\mu_F^2}{m^2}\right)
+(C_F+C_A)^2 \ln^2\left(\frac{\mu_F^2}{m^2}\right) \right. 
\nonumber \\ && \hspace{25mm} \left. 
{}-\zeta_2 (c_3^{gu\rightarrow tg})^2
+\frac{\beta_0}{4} c_3^{gu\rightarrow tg} 
\ln\left(\frac{\mu_R^2}{m^2}\right)\right]
\left[\frac{\ln(s_4/m^2)}{s_4}\right]_+ 
\nonumber \\ && \hspace{-15mm}
{}+\frac{\alpha_s^4(\mu_R)}{\pi^2} 
(-2)(C_F+C_A) \ln\left(\frac{\mu_F^2}{m^2}\right) A^{gu\rightarrow tg} \,  
\left[\frac{\ln(s_4/m^2)}{s_4}\right]_+ 
\nonumber \\ && \hspace{-15mm}
{}+F^B_{gu\rightarrow tg} \frac{\alpha_s^2(\mu_R)}{\pi^2} \left[-(C_F+C_A) c_1^{gu\rightarrow tg} \ln\left(\frac{\mu_F^2}{m^2}\right)
-\zeta_2 c_3^{gu\rightarrow tg} c_2^{gu\rightarrow tg}
+\zeta_3 (c_3^{gu\rightarrow tg})^2 \right.
\nonumber \\ &&  \hspace{15mm}  \left. 
{}-\frac{\beta_0}{4} (C_F+C_A) \ln\left(\frac{\mu_F^2}{m^2}\right) 
\ln\left(\frac{\mu_R^2}{m^2}\right) 
+\frac{\beta_0}{8} (C_F+C_A) \ln^2\left(\frac{\mu_F^2}{m^2}\right) 
\right] \left[\frac{1}{s_4}\right]_+ \, .
\nonumber \\
\eeqa

We note that all the above results also apply to the corresponding process with antiparticles, $g {\bar u} \rightarrow {\bar t} g$.

Finally we consider the massless case, i.e. $m=0$. In that case the 
off-diagonal elements of $\Gamma_S^{(1)}$ can be found from the massive case by simply setting $m=0$. The diagonal elements become
\beqa
\Gamma_{11}^{\rm massless} &=& (C_F+C_A) \ln \left( \frac{-u}{s} \right) 
\nonumber \\
\Gamma_{22}^{\rm massless} &=& \Gamma_{33}^{\rm massless} =  C_F \ln \left( \frac{-u}{s} \right)  
+ \frac{C_A}{2} \ln \left( \frac{-t}{s} \right)
\eeqa
This result is in agreement with that for the process $gq \rightarrow qg$ in Ref. \cite{KOS}, provided one accounts for the known differences between expressions in Feynman gauge (used here) and axial gauge (used in \cite{KOS}). Of course the expressions for the physical NLO and NNLO corrections are the same in both gauges.

\mysection{Numerical results for $gu \rightarrow tg$}

In this section we investigate the numerical effect of the higher-order soft-gluon corrections for the specific Lagrangian of Eq. (\ref{Leff2}).
We begin by presenting the analytical results for the lowest-order hard matrix for this Lagrangian. 
We find
\beq 
H^{(0)} =  \left[\begin{array}{ccc}
H_{11} & N_c H_{11} & H_{13} \\
N_c H_{11} & N^2_c H_{11} & N_c H_{13} \\
H_{13} & N_c H_{13} & H_{33} 
\end{array}\right]
\eeq
where
\beqa
 H_{11} &=&  - g^4_s k_u^2 t \left[2 m^{12}-4 (2 s+t) m^{10}+(2 s+t) (5 s+3 t) m^8 -(s+t) \left(10 s^2+9 t s+t^2\right) m^6 \right.  
\nonumber \\ && \hspace{14mm}\left.
{}+ s (s+t) \left(20 s^2+26 t s+7 t^2\right) m^4 -6 s^2 (s+t)^2 (3 s+2 t) m^2+6 s^3 (s+t)^3  \right] 
\nonumber \\ && 
/ \left[216 m^2 \left(m^2-s\right)^2 s (s+t)^2 \left(-m^2+s+t\right) \right]  \, , 
\eeqa
\beqa
H_{13} &=&  -  g^4_s k_u^2 \left(m^2-2 s-t\right) \left[4 t m^{10}+\left(4 s^2-3 t s-3 t^2\right) m^8+3 (2 s-t) t (s+t) m^6  \right. 
\nonumber \\ && \hspace{38mm} \left. 
{}- 2 s^2 (s+t) (10 s+9 t) m^4 +s^2 (s+t)^2 (24 s+11 t) m^2-8 s^3 (s+t)^3\right]
\nonumber \\ && 
/ \left[ 144 m^2 \left(m^2-s\right)^2 s (s+t)^2 \left(-m^2+s+t\right) \right] \, ,
\eeqa
and
\beqa
H_{33} &=&  - g^4_s k_u^2 \left[2 (2 s+t)^2 m^{12}-\left(44 s^3+66 t s^2+29 t^2 s+3 t^3\right) m^{10} \right.
\nonumber \\ && \hspace{13mm}
{}+ (2 s+t) \left(50 s^3+72 t s^2+25 t^2 s+t^3\right) m^8   
\nonumber \\ && \hspace{13mm}
{}- s (s+t) \left(116 s^3+156 t s^2+60 t^2 s+5 t^3\right)m^6 
\nonumber \\ && \hspace{13mm}
{}+ s (s+t) \left(76 s^4+134 t s^3+67 t^2 s^2+7 t^3 s-t^4\right) m^4 
\nonumber \\ && \hspace{13mm} \left.
{}-s^2 (s+t)^2 \left(32 s^3+42 t s^2+10 t^2 s-t^3\right) m^2+8 s^4 (s+t)^4 \right]
\nonumber \\ && 
/  \left[ 24 m^2 \left(m^2-s\right)^2 s t (s+t)^2 \left(-m^2+s+t\right) \right] \, .
\eeqa

With these results and the previous analytical results in Section 4 we can 
calculate the NLO soft-gluon corrections. We use MSTW2008 parton densities 
\cite{MSTW2008} for our numerical results.

\begin{figure}
\begin{center}
\includegraphics[width=0.6\linewidth]{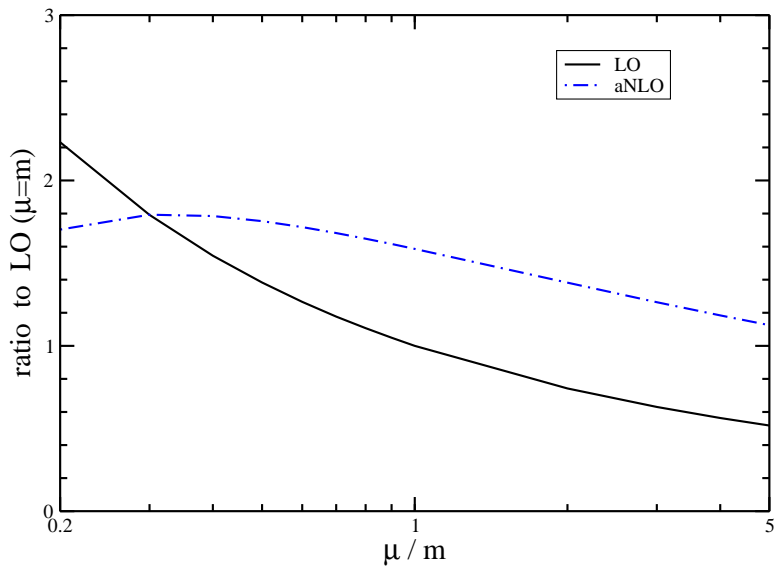}
\caption{The scale dependence for $gu \rightarrow tg$ in $pp$ collisions at 7 TeV LHC energy.}
\label{scale7lhc}
\end{center}
\end{figure}

We first study the effect of the NLO soft-gluon corrections at 7 TeV LHC energy. We find that for the central choice of factorization/renormalization scale, $\mu=m$, the NLO soft-gluon corrections enhance the LO cross section by around 59\%. Thus the inclusion of these corrections is necessary for a more accurate theoretical prediction.  
In Fig. \ref{scale7lhc} we display the scale dependence of the 7 TeV cross section at LO and approximate NLO (aNLO), where aNLO denotes the sum of the LO result and the NLO soft-gluon corrections.
We vary the scale $\mu$ by a factor of five around the central $\mu=m$ value, i.e. between $0.2m$ and $5m$. We see that the scale dependence is reduced at aNLO relative to LO thus providing a more stable theoretical prediction. The usual way to quote the scale uncertainty is to study the variation between $\mu=m/2$ and $2m$. While the uncertainty at LO is 
+38\% -26\%, at NLO it is +11\% -13\%. 

\begin{figure}
\begin{center}
\includegraphics[width=0.6\linewidth]{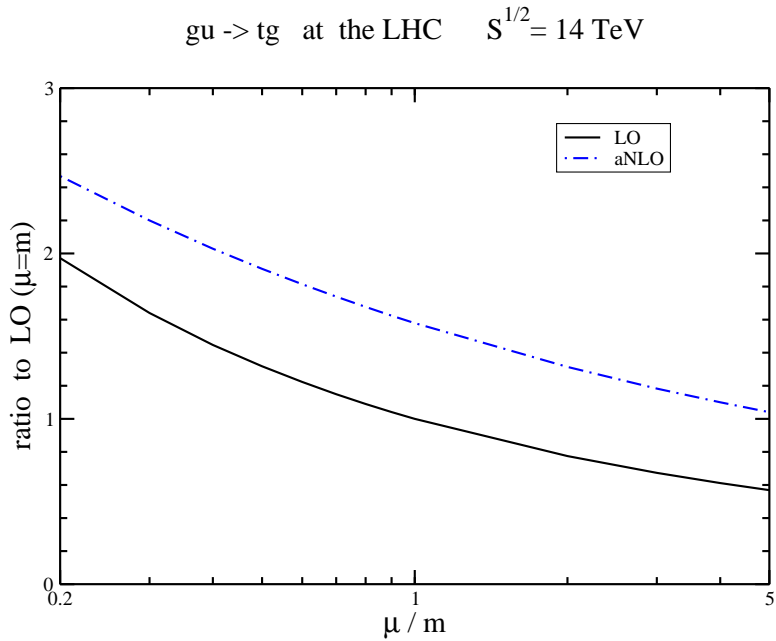}
\caption{The scale dependence for $gu \rightarrow tg$ in $pp$ collisions at 14 TeV LHC energy.}
\label{scale14lhc}
\end{center}
\end{figure}

We next present the corresponding results at 14 TeV LHC energy. 
The NLO soft-gluon corrections at $\mu=m$ enhance the LO cross section by nearly 58\%. In Fig. \ref{scale14lhc} we display the scale dependence of the cross section at LO and aNLO at 14 TeV. While the reduction in scale is not as dramatic
as at 7 TeV, we again see a decrease in the percent uncertainty. While the LO uncertainty with variation over $m/2 \leq \mu \leq 2m$ is +32\% -23\%, at aNLO it is +21\% -17\%. 

As discussed in the previous section, with NLL accuracy we cannot determine fully the coefficients of all powers of the logarithms of the soft-gluon corrections at NNLO; only the two highest powers are fully known. Therefore we do not offer numerical predictions at NNLO.

\mysection{Conclusions}

We have studied FCNC top-quark production via anomalous gluon couplings beyond leading order, in particular the contribution of higher-order corrections to the cross section due to soft-gluon emission. Our work extends previous knowledge on FCNC by considering new couplings and channels.
We have calculated the soft anomalous dimension matrix for top production via anomalous gluon couplings. NLO and NNLO analytical results for the soft-gluon corrections have been derived. We have kept our calculations general so that they can be of wide use and tailored to any specific Lagrangian with anomalous couplings of the gluon to the top quark. These results will enable numerical calculations of higher-order corrections and the setting of new limits in various FCNC models. We have also illustrated the numerical importance of the soft-gluon corrections in a specific model by showing that they significantly enhance the LO cross section while reducing the theoretical uncertainty from scale variation. 

\mysection*{Acknowledgements}

This material is based upon work supported by the National Science Foundation 
under Grant No. PHY 1212472.

\section*{Appendix}

In this Appendix we give some details on the eikonal rules and color 
factor calculations. 

\begin{figure}
\begin{center}
\includegraphics[width=.35\linewidth]{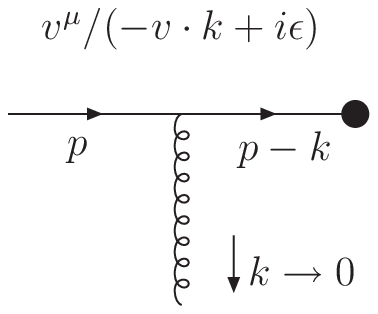}  \hspace{8mm}
\includegraphics[width=.35\linewidth]{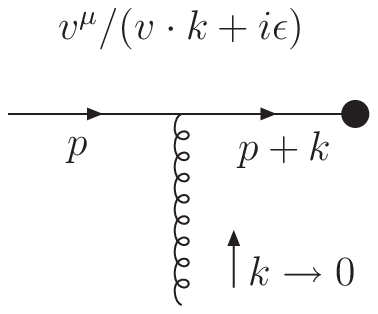}  

\vspace{6mm}

\includegraphics[width=.35\linewidth]{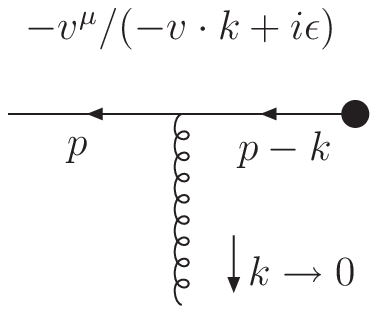}  \hspace{8mm}
\includegraphics[width=.35\linewidth]{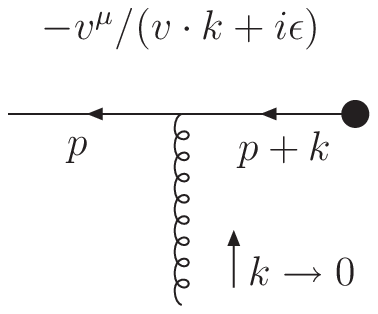}

\vspace{6mm}

\includegraphics[width=.35\linewidth]{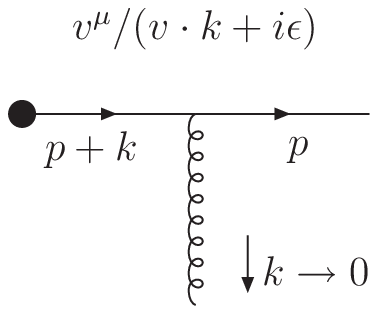}  \hspace{8mm}
\includegraphics[width=.35\linewidth]{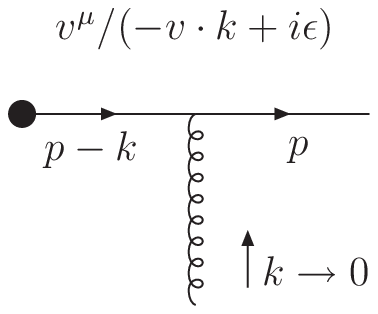}  

\vspace{6mm}

\includegraphics[width=.35\linewidth]{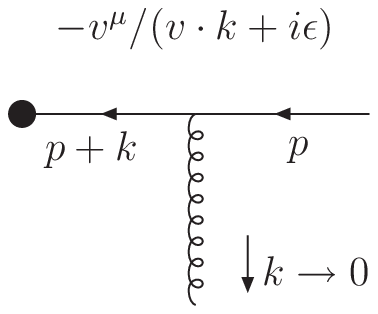}  \hspace{8mm}
\includegraphics[width=.35\linewidth]{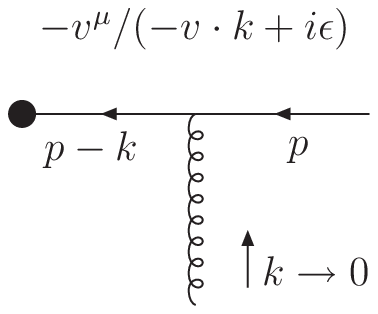}
\caption{Eikonal rules for incoming (top four graphs)
and for outgoing (bottom four graphs) quarks and antiquarks.}
\label{eik-quark}
\end{center}
\end{figure}

\begin{figure}
\begin{center}
\includegraphics[width=.35\linewidth]{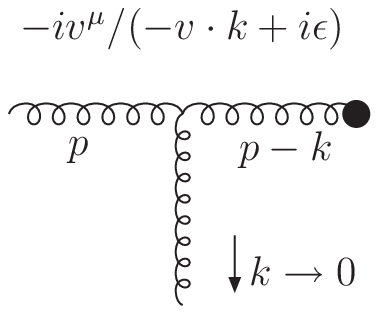}  \hspace{8mm}
\includegraphics[width=.35\linewidth]{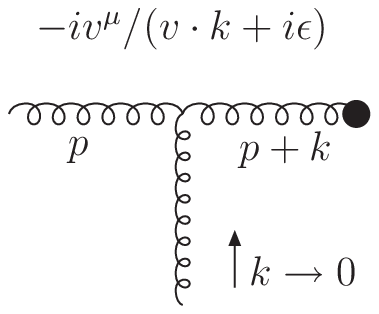}  

\vspace{6mm}

\includegraphics[width=.35\linewidth]{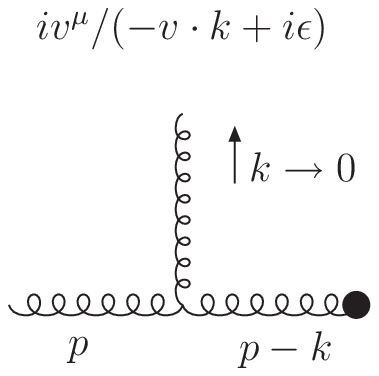}  \hspace{8mm}
\includegraphics[width=.35\linewidth]{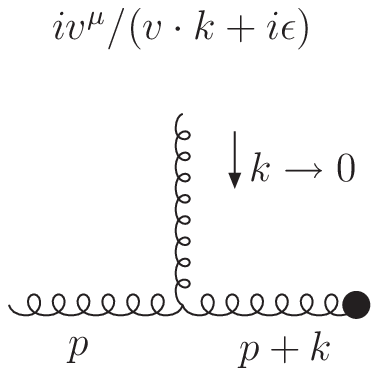}

\vspace{10mm}

\includegraphics[width=.35\linewidth]{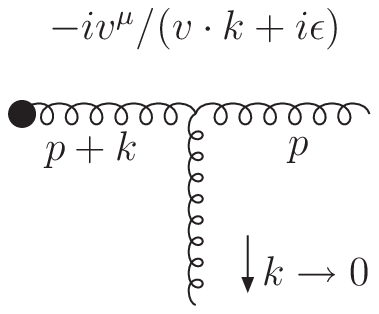}  \hspace{8mm}
\includegraphics[width=.35\linewidth]{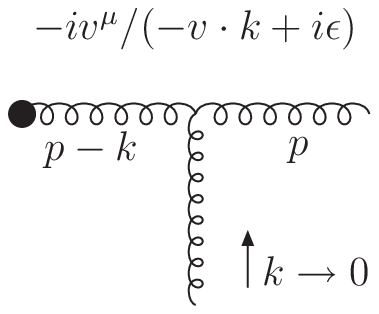}  

\vspace{6mm}

\includegraphics[width=.35\linewidth]{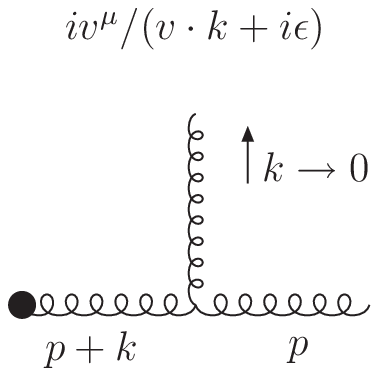}  \hspace{8mm}
\includegraphics[width=.35\linewidth]{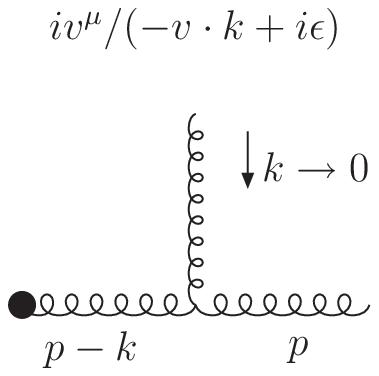}
\caption{Eikonal rules for incoming (top four graphs)
and for outgoing (bottom four graphs) gluons.}
\label{eik-gluon}
\end{center}
\end{figure}

We begin with the eikonal rules. They are shown in Fig. \ref{eik-quark} for 
quark and antiquark eikonal lines and in Fig. \ref{eik-gluon} for gluon 
eikonal lines. The expression for each rule is given above the corresponding graph in these figures.

In the eikonal approximation the usual Feynman rules are simplified by letting the gluon momentum  approach zero. For example, for an outgoing quark line
that emits a gluon of momentum $k$ and has final momentum $p$, the usual
Feynman rule in terms of the Dirac spinor, vertex factors, and propagator for 
the internal line reduces to the eikonal rule as follows:
\beqa
{\bar u}(p) \, (-i g_s T^c) \, \gamma^{\mu}
\frac{i (p\!\!/+k\!\!/+m)}{(p+k)^2
-m^2+i\epsilon} \rightarrow {\bar u}(p)\,  g_s T^c \, \gamma^{\mu}
\frac{p\!\!/+m}{2p\cdot k+i\epsilon}
={\bar u}(p)\, g_s T^c \,
\frac{v^{\mu}}{v\cdot k+i\epsilon}
\nonumber
\eeqa

In the rules shown in Figs. \ref{eik-quark}, \ref{eik-gluon} we do not show the 
color-factor part, which
is $T^c$ for quark lines and $f^{abc}$ for gluon lines.  These color factors 
are calculated separately as shown next.

\vspace{2mm}

The calculation of color factors proceeds by weighting the vertex with a given basis and  contracting the color factors given by the Feynman rules. Each color factor is expressed in its final form as a linear combination of the basis tensors.
The color calculations are for color indices as defined in 
Eq. (\ref{colorbasis}),
and indices corresponding to the interior of loops are primed. 
We use standard color identities and show our results below.

\vspace{3mm}

{\bf Color factors for $c_1 = \delta_{b i} \delta_{a j}$}

\vspace{2mm}

First we calculate the color factors corresponding to the first color tensor in the chosen basis, $ \delta_{b i} \delta_{a j}$.

\vspace{1mm}

For diagram 1: $f^{a' a c} T^c_{b' b} \delta_{b' i} \delta_{a' j} = i \times c_3 $

For diagram 2: $T^c_{i i'} f^{j c j'}  \delta_{b i'} \delta_{a j'} = -i \times c_3$

For diagram 3: $f^{c a a'} T^c_{i i'}  \delta_{b i'} \delta_{a' j}  = -i \times c_3$   

For diagram 4: $T^c_{b' b} f^{j j' c}  \delta_{b 'i} \delta_{a j'} =  i \times c_3$

For diagram 5: $T^c_{i i'} T^c_{ b' b } \delta_{b' i'} \delta_{a j} = C_F \times c_1$

For diagram 6: $f^{j  c j'} f^{a' c a} \delta_{b i} \delta_{a' j'} = - C_A \times c_1$ 

\vspace{3mm}

{\bf Color factors for $c_2 = d^{a j c} T^c_{i b}$}

\vspace{2mm}

Next we calculate the color factors corresponding to the second color tensor in the chosen basis, $ d^{a j c} T^c_{i b}$.

\vspace{1mm}

For diagram 1:
$f^{a' a c} T^c_{b' b} d^{a' j d} T^d_{i b'} = -i \frac{N_c}{4} \times c_2 +i \left( \frac{N_c}{4} - \frac{1}{N_c} \right) \times c_3$

For diagram 2:
$T^c_{i i'} f^{j c j'}  d^{a j' d} T^d_{i' b}  =  i \frac{N_c}{4} \times c_2 - i \left( \frac{N_c}{4} - \frac{1}{N_c} \right) \times c_3$

For diagram 3:
$f^{ a a' c } T^c_{i i'}    d^{a' j d} T^d_{i' b}  =  - i \frac{N_c}{4} \times c_2 - i \left( \frac{N_c}{4} - \frac{1}{N_c} \right) \times c_3 $

For diagram 4:
$f^{j j' c} T^c_{b' b}   d^{a j' d} T^d_{i b'} =  i \frac{N_c}{4} \times c_2 +i \left( \frac{N_c}{4} - \frac{1}{N_c} \right) \times c_3 $

For diagram 5:
$T^c_{i i'} T^c_{b' b }   d^{a j d}  T_{i' b'}^d =   \left( C_F - \frac{C_A}{2} \right) \times c_2$

For diagram 6:
$f^{a  a' c} f^{j c j'}   d^{a' j' d} T^d_{i b} = -\frac{C_A}{2} \times c_2 $

\vspace{3mm}

{\bf Color factors for $c_3 = i  f^{a j c} T^c_{i b}$}

\vspace{2mm}

Finally we calculate the color factors corresponding to the third color tensor in the chosen basis, $ i  f^{a j c} T^c_{i b}$.

\vspace{1mm}

For diagram 1: $f^{a' a c} T^c_{b' b} i f^{a' j d} T^d_{i b'}  = \frac{i}{2} \times c_1  + i \frac{N_c}{4} \times c_2 -i \frac{N_c}{4} \times c_3 $

For diagram 2:
$f^{j c j'} T^c_{i i'} i f^{a j' d} T^d_{i'b}  = - \frac{i}{2}\times c_1- i \frac{N_c}{4} \times c_2 +i \frac{N_c}{4} \times c_3  $

For diagram 3:
$f^{a a' c} T^c_{i i'} i f^{a' j d} T^d_{i'b}  = - \frac{i}{2}\times c_1- i \frac{N_c}{4} \times c_2 -i \frac{N_c}{4} \times c_3  $

For diagram 4:
$f^{j  j' c} T^c_{b' b} i f^{a j' d} T^d_{ib'}  = \frac{i}{2} \times c_1  + i \frac{N_c}{4} \times c_2 + i \frac{N_c}{4} \times c_3 $

For diagram 5:
$T^c_{ i i'} T^c_{b' b } i  f^{a j d} T_{i' b'}^d = \left( C_F - \frac{C_A}{2} \right) \times c_3 $

For diagram 6:
$f^{a'  c a} f^{j c j'}  i  f^{a' j' d} T_{i b}^d = - \frac{C_A}{2} \times c_3 $


\begin{thebibliography}{99}

\bibitem{MT}	
E. Malkawi and T. Tait, Phys. Rev. D {\bf 54}, 5758 (1996) [hep-ph/9511337].

\bibitem{HWYZ}
T. Han, K. Whisnant, B.L. Young, and X. Zhang, Phys. Lett. B {\bf 385}, 311 
(1996) [hep-ph/9606231]. 

\bibitem{TY96}
T. Tait and C.P. Yuan, Phys. Rev. {\bf D 55}, 7300 (1997) [hep-ph/9611244];
\newline
Phys. Rev. D {\bf 63}, 014018 (2000) [hep-ph/0007298]. 

\bibitem{HHWYZ}
M. Hosch, K. Whisnant, and B.L. Young,  Phys. Rev. D {\bf 56}, 5725 (1997) 
[hep-ph/9703450];  \newline
T. Han, M. Hosch, K. Whisnant, B.L. Young, and X. Zhang, 
Phys. Rev. D {\bf 58}, 073008 (1998) [hep-ph/9806486].

\bibitem{ABNK}
A. Belyaev and N. Kidonakis, Phys. Rev. D {\bf 65}, 037501 (2002) 
[hep-ph/0102072].

\bibitem{NKAB}
N. Kidonakis and A. Belyaev, JHEP {\bf 0312} (2003) 004 [hep-ph/0310299]. 

\bibitem{NKEM}
N. Kidonakis and E. Martin, in Proceedings of DPF 2013, arXiv:1310.0363 [hep-ph].

\bibitem{ATLAS1}
ATLAS Collaboration, G. Aad {\it et al.}, Phys. Lett. B {\bf 712}, 351 (2012) [arXiv:1203.0529 [hep-ex]].

\bibitem{ATLAS2}
ATLAS Collaboration, ATLAS-CONF-2013-063.

\bibitem{GSW}
R. Guedes, R. Santos, and M. Won, Phys. Rev. D {\bf 88}, 114011 (2013)
[arXiv:1308.4723 [hep-ph]].

\bibitem{NKGS}
N. Kidonakis and G. Sterman, Phys. Lett. B {\bf 387}, 867 (1996);
\newline
Nucl. Phys. B {\bf 505}, 321 (1997) [hep-ph/9705234].

\bibitem{KOS}
N. Kidonakis, G. Oderda, and G. Sterman, Nucl. Phys. B {\bf 531}, 365 (1998) [hep-ph/9803241]. 

\bibitem{GS87}
G. Sterman, Nucl. Phys. B {\bf 281}, 310 (1987).

\bibitem{CT89}
S. Catani and L. Trentadue, Nucl. Phys. B {\bf 327}, 323 (1989).

\bibitem{NKtW}
N. Kidonakis, Phys. Rev. D {\bf 82}, 054018 (2010) [arXiv:1005.4451 [hep-ph]].

\bibitem{MSTW2008}
A.D. Martin, W.J. Stirling, R.S. Thorne, and G. Watt, 
Eur. Phys. J. C {\bf 63}, 189 (2009) [arXiv:0901.0002 [hep-ph]].

\end{thebibliography}
\end{document}